\documentclass[prl, aps, 12pt, showkeys, showpacs,tightenlines] {revtex4}
\usepackage{amsmath}
\usepackage{epsfig}
\usepackage{hyperref}
\begin{document}
\title{Entropy and Ionic Conductivity}
\author{Yong-Jun Zhang}
\email{yong.j.zhang@gmail.com}
\affiliation{Science College, Liaoning Technical University, Fuxin, Liaoning 123000, China}

\begin{abstract}
It is known that the ionic conductivity can be obtained by using the diffusion constant and the Einstein relation. We derive it here by extracting it from the steady electric current which we calculate in three ways, using statistics analysis, an entropy method, and an entropy production approach.
\end{abstract}
\keywords{ionic conductivity; fluctuation; entropy; entropy production}
\pacs{72.80.-r, 05.60.Cd, 05.70.Ln, 05.40.-a}
\maketitle

\section{introduction}
Electric conductivity, $\sigma$, is a coefficient that appears in Ohm's law,
\begin{equation}\label{jv}
	j=\sigma E,
\end{equation}
where $j$ is the electric current density and $E$ is the electric field. For ionic conduction, $\sigma$ is also called the ionic conductivity. The ionic conductivity can be easily derived by studying
\begin{figure}[htbp]
  \begin{center}
    \mbox{\epsfxsize=5.0cm\epsfysize=5.0cm\epsffile{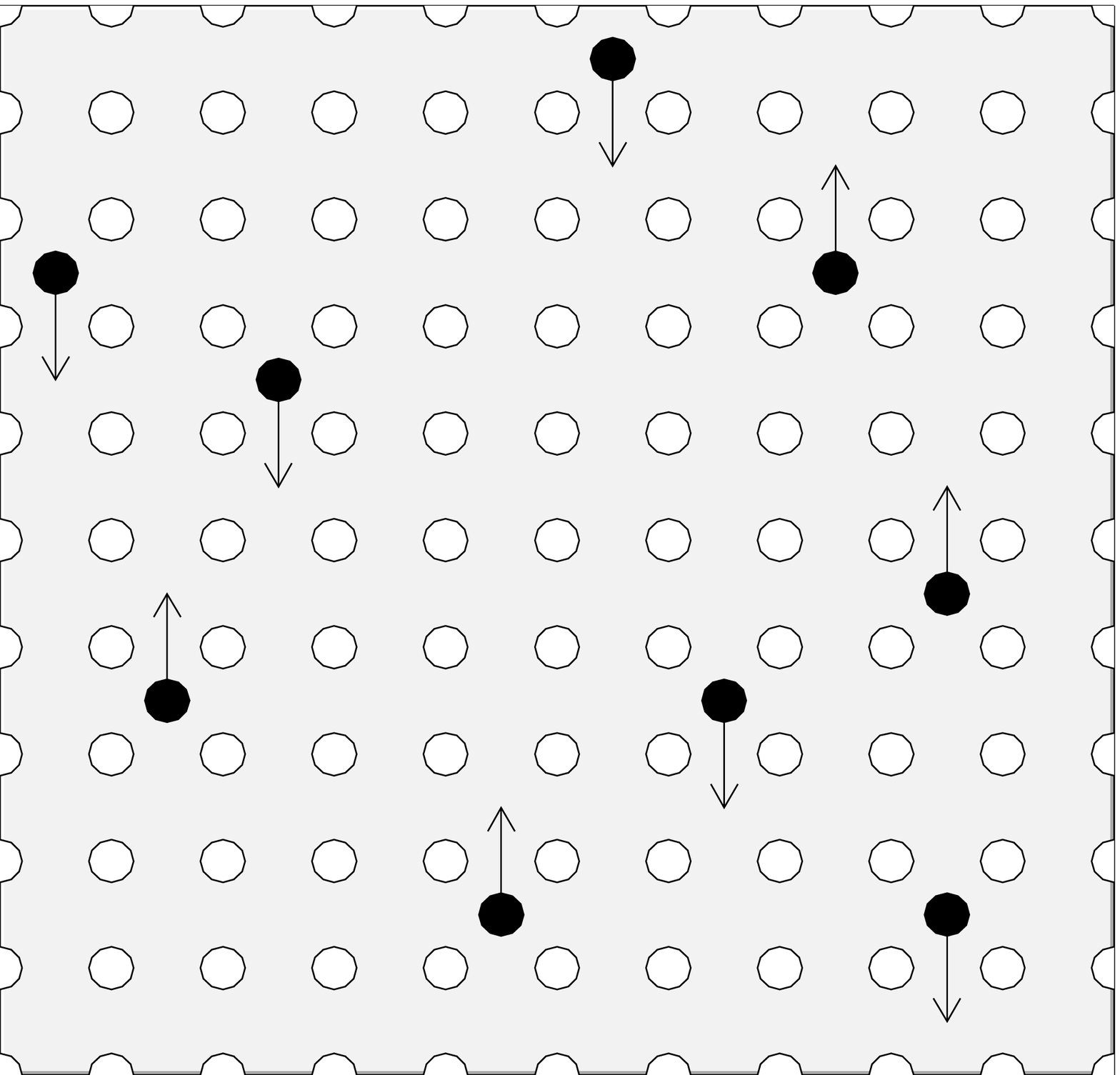}
        \ \ \ \ \ \epsfxsize=5.0cm\epsfysize=5.0cm\epsffile{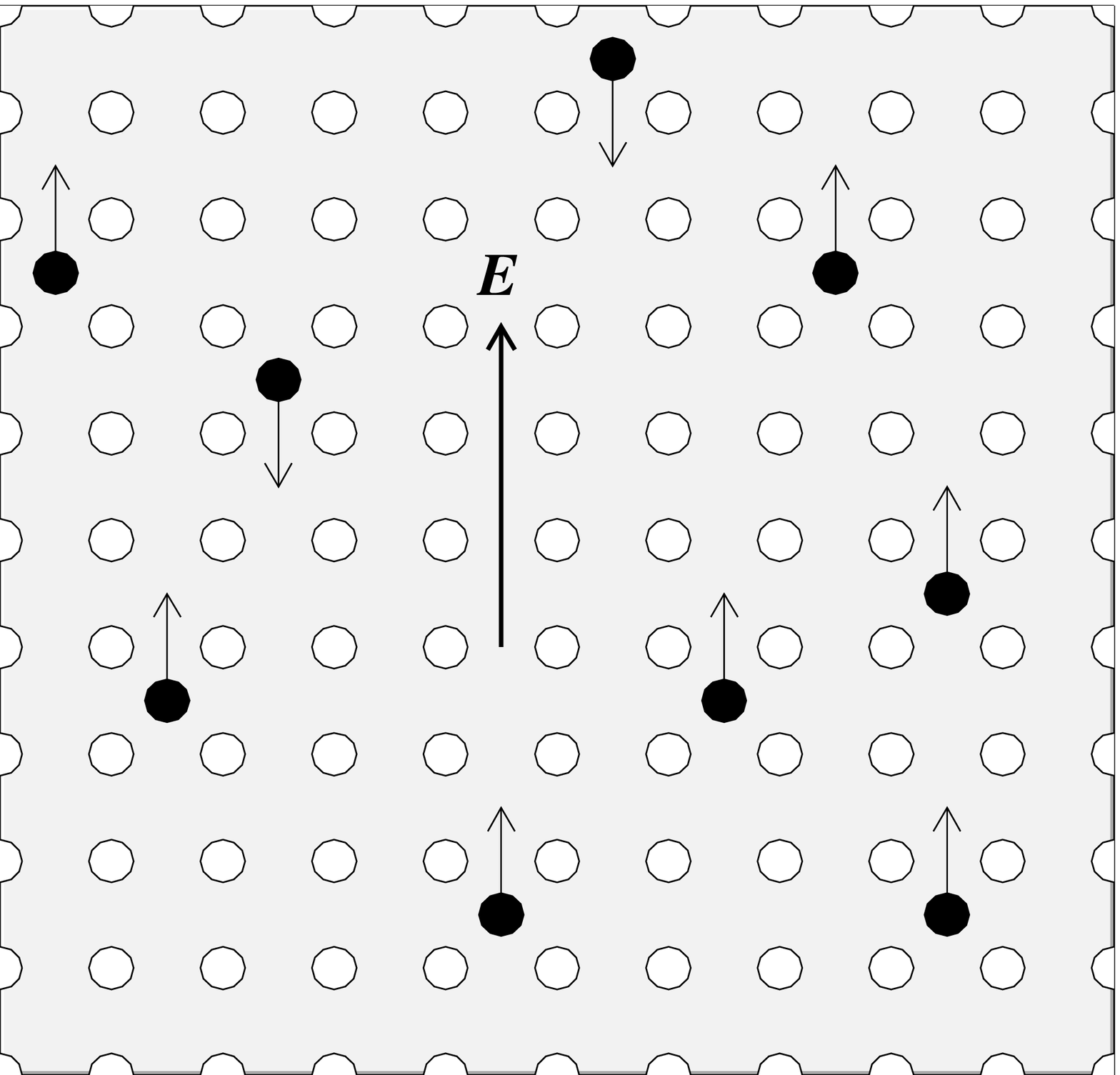}
	}
  \end{center}
\caption{
A simple ionic conductor. The white circles are the non-movable lattice, and the black circles are the movable interstitial ions that can jump from site to site. For the right graph, there is an external electric field pointing upward. The electric field will affect the jump of ions along its direction. So one only needs to study how ions jump up and down. 
\label{ions}}
\end{figure}
an ionic conductor that is composed simply of a non-movable crystal lattice and movable interstitial ions. The ions can jump from site to site, as shown in Fig. \ref{ions}, and external electric field $E$ can affect their jumps. If $E$ is chosen pointing upward, one only needs to study how an ion jumps up and down. 

Given the ionic jump frequency
\begin{equation} 
	\nu\exp\left(-\frac{\varepsilon}{k_BT}\right),
\end{equation}
one can get from kinetic theory the diffusion constant \cite{Kittel} as
\begin{equation} 
	D=a^2\nu\exp\left(-\frac{\varepsilon}{k_BT}\right),
\end{equation}
where $\nu$ is an effective vibration frequency, $\varepsilon$ is the height of the barrier which an ion must surmount in order to pass to the adjacent site, $a$ is the lattice constant, $k_B$ is the Boltzmann constant and $T$ is the temperature.
The diffusion constant is related to the ion mobility $\mu$ by the Einstein relation,
\begin{equation} 
	\mu=\frac{q}{k_BT}D,
\end{equation}
where $q$ is the interstitial ion charge. Thus one gets the ionic conductivity \cite{Kittel} as
\begin{equation} \label{kd}
	\sigma=nq\mu=\frac{na^2q^2}{k_BT}\nu\exp\left(-\frac{\varepsilon}{k_BT}\right),
\end{equation}
where $n$ is the ion concentration. One may also obtain the same ionic conductivity by calculating the ion drift velocity,
\begin{equation} \label{v_d}
	v_d=a\left[\nu\exp\left(-\frac{\varepsilon-E qa/2}{k_BT}\right)-\nu\exp\left(-\frac{\varepsilon+E qa/2}{k_BT}\right)\right]\approx\nu\exp\left(-\frac{\varepsilon}{k_BT}\right)\frac{a^2q}{k_BT}E.
\end{equation}
With the ion drift velocity, one can obtain the electric current density,
\begin{equation} 
	j=nqv_d=\frac{na^2q^2}{ k_BT}\nu\exp\left(-\frac{\varepsilon}{k_BT}\right)E,
\end{equation}
from which one can extract again the same ionic conductivity Eq. (\ref{kd}).

We shall derive the ionic conductivity again in three different ways, using statistics analysis, an entropy method, and an entropy production approach. All our derivations are subject to three constraints.
\begin{enumerate}
\setlength{\itemsep}{-3pt}
\item $N\gg 1$: the conductor is large enough to contain a large number ($N$) of ions. 
\item $n\sim 0$: the ions are dilute, so each ion jumps independently.
\item $E\sim 0$: the external electric field is weak, so the electric current is weak, $I\sim 0$.
\end{enumerate}

\section{statistics analysis}
From Fig. \ref{ions}, we understand that when the ion conductor is of small size ( this is the same as saying that $N$ is small ), the electric current will fluctuate. But when the conductor is of large size ( i.e., $N\gg 1$ ), the electric current will become steady. The steady electric current is thus actually the most probable electric current, which we will find in this section by using statistics analysis. Then we can extract the ionic conductivity.

For the ionic conductor as shown in Fig. \ref{ions}, we separate the system from the environment. The system only consists of all the interstitial ions (and only about how they jump). The environment includes everything else. Since we choose $n\sim 0$, the ions are far from each other. Thus, when they jump, they do not affect each other. So we may imagine that all the ions jump in a synchronized way and that each jump takes the same time $\tau$. This allows us to easily study what the jump configuration is and what the corresponding probability is.

First, as a start, let the system consist of only one ion. That one ion, in time $\tau$, either jumps up or jumps down. So we have
\begin{equation} \label{1ion}
\begin{array}{c||c|c||l}\hline
	{\rm configuration} 	&\uparrow	&\downarrow &\\ \hline
	\Omega_S &1&1 \\
	I_V&aq/\tau& -aq/\tau & I_V=\sum\limits_{i=1}^Nqv_i\\
	 Q &aqE&-aqE & Q=I_VE\tau \\
	\Delta S_E&\frac{aqE}{T}&-\frac{aqE}{T}& \Delta S_E=\frac{Q}{T}\\
	\Omega_E &\propto\exp(\frac{aqE}{k_BT})& \propto\exp(-\frac{aqE}{k_BT})&\Omega_E\propto\exp\left(\frac{\Delta S_E}{k_B}\right)\\
	P & \propto 1\times \exp(\frac{aqE}{k_BT})& \propto 1\times \exp(-\frac{aqE}{k_BT})& P\propto \Omega_E\Omega_S\\\hline
\end{array}
\end{equation}
where
\begin{itemize}\setlength{\itemsep}{-3pt}
\item $\uparrow$ represents the state transition in which the ion jumps up; 
\item $\downarrow$ represents the state transition in which the ion jumps down; 
\item $\Omega_S$ is the number of system microscopic states from which the transition can occur;
\item $I_V=jV=\sum\limits_{i=1}^Nqv_i$, where $V$ is the conductor volume and $v=\pm a/\tau$, the effective ion velocity; 
\item $Q$ is the heat exchanged with the environment;
\item $\Delta S_E$ is the entropy change of the environment during $\tau$;
\item $\Omega_E$ is the number of environment microscopic states compatible with the transition;
\item $P$ is the probability for an $I_V$ to occur, and for one ion system $P$ to become the Boltzmann distribution; but this $P$ is different from its counterpart in Eq. (\ref{v_d}) by an extra factor 2 which will persist to the end. 
\end{itemize} 
%

Then, let the system consist of two ions. Eq. (\ref{1ion}) becomes 
\begin{equation} \label{2ions}
\begin{array}{c||c|c|c}\hline
	{\rm configuration}	&\uparrow\uparrow	&\uparrow \downarrow {\rm \ or\ } \downarrow\uparrow&\downarrow\downarrow\\\hline
	\Omega_S&1&2&1\\
        I_V&2aq/\tau&0& -2aq/\tau\\
	Q &2aqE&0&-2aqE\\
	\Delta S_E&\frac{2aqE}{T}&0&-\frac{2aqE}{T}\\
	\Omega_E &\propto\exp(\frac{2aqE}{k_BT})&\propto1&\propto\exp(-\frac{2aqE}{k_BT})\\
	P&\propto1\times \exp(\frac{2aqE}{k_BT})&\propto2\times 1&\propto1\times \exp(-\frac{2aqE}{k_BT})\\\hline
\end{array} 
\end{equation}

Finally, let the system consist of $N$ ions. Eq. (\ref{1ion}) becomes
\begin{equation} \label{Nions}
\begin{array}{c||ccccc}\hline
	\Omega_S &1&\cdots&C_N^k&\cdots&1\\
        I_V&N{aq}/{\tau}&\cdots&(2k-N)aq/\tau&\cdots& -Naq/\tau\\
	Q&NaqE&\cdots&(2k-N)aqE&\cdots&-NaqE\\
	\Delta S_E & \frac{NaqE}{T} &\cdots &\frac{(2k-N)aqE}{T}& \cdots & -\frac{NaqE}{T}\\
	\Omega_E & \propto \exp\left(\frac{NaqE}{k_BT}\right) &\cdots &\propto\exp\left(\frac{(2k-N)aqE}{k_BT}\right)& \cdots & \propto \exp\left(-\frac{NaqE}{k_BT}\right)\\
	P&\propto 1\times \exp\left({\frac{NaqE}{k_BT}}\right)&\cdots&\propto C_N^k\times \exp\left({\frac{(2k-N)aqE}{k_BT}}\right)&\cdots&\propto 1\times \exp\left(-\frac{NaqE}{k_BT}\right)\\\hline
\end{array} 
\end{equation}
where $k$ is the number of ions that jump up. The most probable $k$ can be obtained by
\begin{equation}\label{variation} 
	\frac{\delta P}{\delta k}=0,
\end{equation}
or by
\begin{equation} \label{delta_ln_P}
	\frac{\delta \ln P}{\delta k}=0,
\end{equation}
which becomes 
\begin{equation} 
	 \frac{\delta\ln C_N^k}{\delta k}+\frac{2aqE}{k_BT}=0.
\end{equation}
Since we choose $E\sim 0$, we have $k\sim N/2$; and since we choose $N\gg 1$, we can apply Stirling's formula. So we can write 
\begin{equation}\label{Stirling}
	\ln C_N^k\approx \ln C_N^{\frac{N}{2}}-\frac{2(k-\frac{N}{2})^2}{N}.
\end{equation}
Thus we obtain the most probable $k$ as
\begin{equation} \label{most_probable_k}
	k=\frac{NaqE}{2k_BT}+\frac{N}{2}.
\end{equation}
Then the most probable $I_V$ is
\begin{equation} \label{I_V}
	I_V=\frac{Na^2q^2E}{k_BT\tau}.
\end{equation}
This is also the steady $I_V$. 
So the steady electric current density is
\begin{equation} 
	j=\frac{I_V}{V}=\frac{na^2q^2}{k_BT\tau }E,
\end{equation}
and the ionic conductivity is extracted as
\begin{equation}\label{sigma_statistics} 
	\sigma=\frac{na^2q^2}{k_BT\tau }.
\end{equation}
The ion jump time relates to the ion jump frequency. Since in this study we only consider how an ion jumps up or down, we should have
\begin{equation} 
	\tau=\frac{1}{2\nu\exp(-\frac{\varepsilon}{k_BT})}.
\end{equation}
So, Eq. (\ref{sigma_statistics}) is the same as Eq. (\ref{kd}), except that there is an extra factor 2.

In our study, we have assumed that all ions are synchronized to jump, and that each jump takes the same time interval. In reality, each ion jumps randomly and independently. Though the pictures are different, the result is the same.

\section{entropy method}
In this section, we shall find the most probable electric current again, but with an entropy method. Here the entropy consists of two parts: the environment entropy and the system entropy. The environment entropy is obtained from Eq. (\ref{Nions}) as
\begin{equation} \label{S_E}
	S_E=S_{E0}+\Delta S_E=S_{E0}+\frac{(2k-N)aqE}{T},
\end{equation}
where $S_{E0}$ is the initial environment entropy when the ions are not yet ready to jump. Using Eq. (\ref{Stirling}), the system entropy is
\begin{equation} \label{S_S}
	S_S=k_B\ln\Omega_S=k_B\ln C_N^k\approx S_{S0}-\frac{2k_B(k-\frac{N}{2})^2}{N},
\end{equation}
where $S_{S0}$ is the system equilibrium entropy.
The total entropy is
\begin{equation} 
	S=S_E+S_S.
\end{equation}
Then the most probable $k$ can be obtained by using
\begin{equation} 
	\frac{\delta S}{\delta k}=0
\end{equation}
which is equivalent to Eq. (\ref{variation}).

We can also directly calculate the most probable electric current by using
\begin{equation} 
	\frac{\delta S}{\delta I}=0.
\end{equation}
To do that, we need write $S$ as a function of $I$. Using Eq. (\ref{Nions}),
\begin{equation} 
	I=jA=\frac{I_V}{V}A=\frac{I_V}{h}=\frac{(2k-N)aq}{h\tau},
\end{equation}
and using Eq. (\ref{S_S}), the system entropy is rewritten as
\begin{equation} \label{S_SI}
	S_S=S_{S0}-\frac{1}{2}\frac{k_Bh^2\tau^2}{Na^2q^2}I^2,
\end{equation}
which takes the same form as the entropy of a dilute gas that carries a heat flux \cite{yjzhang} or a velocity gradient \cite{yjzhang2}.
Using Eq. (\ref{S_E}) the environment entropy is rewritten as
\begin{equation} 
	S_E=S_{E0}+\frac{Eh\tau}{T}I.
\end{equation}

\begin{figure}[htbp]
  \begin{center}
    \mbox{\epsfxsize=12.0cm\epsfysize=8.0cm\epsffile{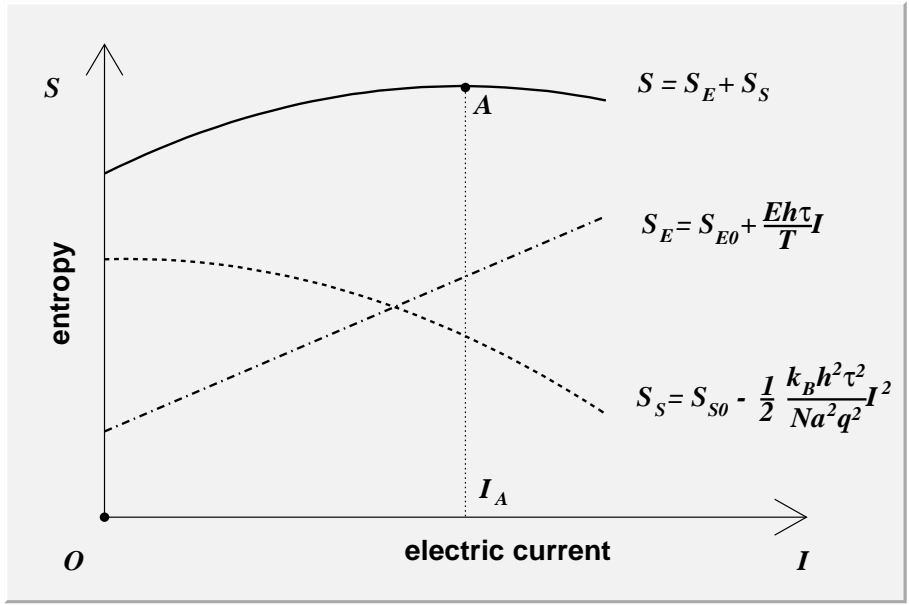}}
  \end{center}
\caption{
The entropies. 
The dashed line is the system entropy. The dash-dotted line is the environment entropy. The solid line is the total entropy. The total entropy has its maximum value at point A. $I_A$ is the most probable electric current, which is also the steady electric current for a large enough conductor.
\label{fig2}}
\end{figure}

All the entropies are shown in Fig. \ref{fig2}. Only at the most probable electric current does the total entropy have its maximum value. The most probable electric current is also the steady electric current. Thus the steady electric current can be obtained by maximizing the entropy. It is obtained as
\begin{equation} \label{themostprobableI}
	I=\frac{Na^2q^2}{hk_BT\tau}E.
\end{equation}
The corresponding steady electric current density is then
\begin{equation} 
	j=\frac{I}{A}=\frac{na^2q^2}{k_BT\tau}E,
\end{equation}
from which the ionic conductivity Eq. (\ref{sigma_statistics}) is extracted again.

In addition, Fig. \ref{fig2} indicates that the system entropy and the environment entropy compete with each other, and their competition determines the value of the steady electric current. 
Entropy always tends to increase. When the environment entropy tends to increase, it drives the electric current $I$ to increase too. When the system entropy tends to increase, it drives $I$ to decrease in the opposite direction. Only at the steady $I$ do the two entropies balance each other, and at the same time, the total entropy has its maximum value. Similar discussions can also be found in \cite{yjzhang3, yjzhang4}.

\section{entropy production approach}
In this section, we shall see if the steady electric current can be obtained by another approach that uses the entropy production. We shall discuss this by analogy with studies of the thermal conductivity \cite{yjzhang} and viscosity \cite{yjzhang2} of a dilute gas.

For a given electric current $I$, when it is steady, the entropy production is
\begin{equation}\label{EhI_T} 
	\sigma_I=\frac{Eh}{T}I,
\end{equation}
where the subscript $I$ in symbol $\sigma_I$ stands for electric current.
When the system relaxes from an unsteady state, the entropy production is different. And the shorter the relaxation time, the larger the entropy production. If the relaxation is exponential, and if the minimum relaxation time is the jump time $\tau$, one may write the electric current that relaxes at the maximum rate as
\begin{equation} \label{I_R}
	I_R(t)=I\exp\left(-\frac{t}{\tau}\right).
\end{equation}
Using Eq. (\ref{S_SI}), the system entropy during relaxation is 
\begin{equation} 
	S_S(t)=S_{S0}-\frac{1}{2}\frac{k_Bh^2\tau^2}{Na^2q^2}I^2\exp\left(-\frac{2t}{\tau}\right),
\end{equation}
and the entropy production in the limit $t\to 0$ is 
\begin{equation} \label{sigma_I}
	\sigma_I=\left.\frac{dS(t)}{dt}\right |_{t=0}=\frac{k_Bh^2\tau}{Na^2q^2}I^2.
\end{equation}

\begin{figure}[htbp]
  \begin{center}
    \mbox{\epsfxsize=12.0cm\epsfysize=8.0cm\epsffile{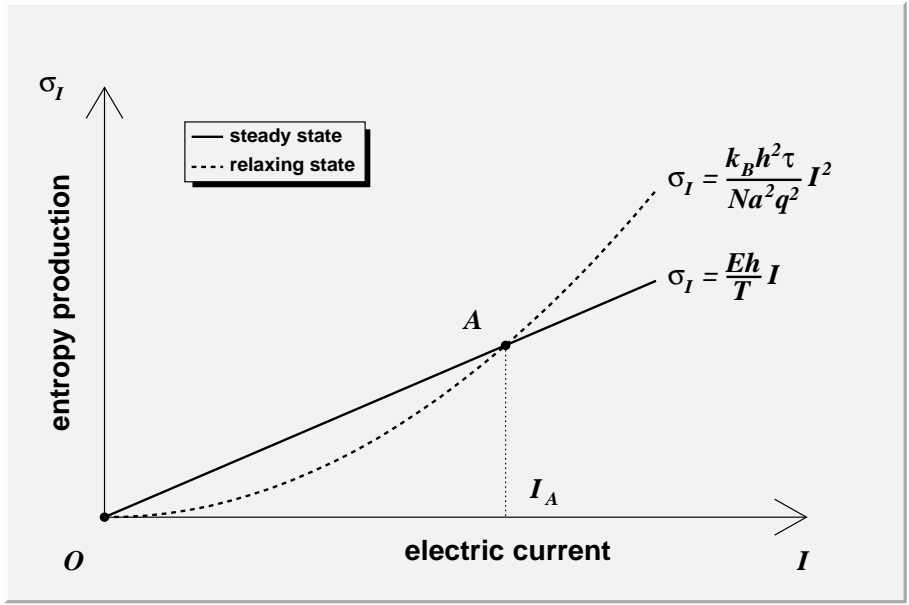}}
  \end{center}
\caption{
Entropy production with respect to electric current. For the solid line, the electric current is steady. For the dashed line, the electric current is about to start to relax at the maximum rate. The steady electric current is $I_A$. For an electric current $I>I_A$, it cannot be steady, because in this range the entropy production of the steady state is at least lower than that of a relaxing state. The steepest entropy ascent (SEA) principle indicates that a state can occur only if it has the maximum entropy production. Thus the steady electric current must be in the range $0\le I\le I_A$. Subsequently, according to the maximum entropy production (MEP) principle, the steady electric current must be $I_A$, because among all candidates it has the maximum entropy production.
\label{entropy_production}}
\end{figure}

Relations (\ref{EhI_T}) and (\ref{sigma_I}) between the entropy production and the current for a steady and a relaxing state are shown in Fig. \ref{entropy_production}. By the steepest entropy ascent (SEA) principle \cite{Beretta3,Beretta32,Beretta4,Beretta5,Beretta6}, we know that an electric current $I$ bigger than $I_A$ cannot be steady, because in that range the steady electric current does not have the steepest entropy ascent. Thus the steady electric current must be in the range $[0,I_A]$. Then the maximum entropy production (MEP) principle \cite{Paltridge1, Paltridge2, Paltridge3,Ziegler,MEPP,Dewar1,Dewar2,Niven} indicates that the steady electric current must be $I_A$, because among all candidates $I_A$ has the maximum entropy production. Thus the steady electric current is obtained from $I_A$ as
\begin{equation} 
	I=\frac{Na^2q^2}{hk_BT\tau}E,
\end{equation}
which is the same as Eq. (\ref{themostprobableI}). So the same ionic conductivity \ref{sigma_statistics} is reproduced again.

\section{Conclusion}
We derive the ionic conductivity by extracting it from the steady electric current, which we obtain by using three different approaches. ({\bf I}) The first approach is a statistics analysis, which is based on calculating the most probable electric current. For a sufficiently large ionic conductor, the electric current can be steady, and it is actually the most probable electric current, which can be obtained by statistics analysis. 
({\bf II}) The second approach is an entropy method. In this approach, we calculate the system entropy and the environment entropy, and study their competition. The system entropy is related to the way in which each ion moves. The environment entropy is related to the heat generated by the electric current. Since both entropies tend to increase, they compete with each other to drive the electric current in opposite directions. When they balance each other, the electric current becomes steady and the total entropy is maximum. 
({\bf III}) The last approach uses entropy production principles. In this approach, one needs to assumes that, for a given electric current, its relaxation is exponential in time and that the minimum relaxation time is the ion jump time. Then, by using two entropy production principles, the SEA principle and the MEP principle, one can obtain the steady electric current. 

All three approaches produce the same steady electric current. And so the same ionic conductivity is extracted, which is also the same as the kinetic theory result, but with an extra factor 2.

\end{document}